\documentstyle[12pt,epsf]{article}
\def\vday{{Oct13}}
\def\chk{
$\Rightarrow$\marginpar{$\Leftarrow$\tiny{\vday}} 
}

\newcommand{\sss}{\scriptscriptstyle}
\newcommand{\VKM}{V_{\rm \sss KM}}
\def\betab{\mbox{\boldmath $\beta$}}
\def\crb{\mbox{\boldmath $C_r$}}

\def\CA{{{N^2-1\over 2N}}}

\begin{document}

\thispagestyle{empty} \null\vskip -1cm 
\centerline{\hfill
FERMILAB-Pub-98/366-T}
\centerline{hep-ph/9811354\hfill
NHCU-HEP-97-02}
\vskip 2cm
\begin{center}
{\large {\bf Vector Quark Model and 
$B \rightarrow X_s \gamma$ Decay}} \vspace{1cm} 
\end{center}
\centerline   
{Chia-Hung V. Chang$^{(1,2)}$ Darwin Chang$^{(1,3)}$, and Wai-Yee
Keung$^{(4,3)}$ }

\begin{center}{\it
$^{(1)}$NCTS and Physics Department, National Tsing-Hua University,\\
Hsinchu, 30043, Taiwan, R.O.C.\\[0pt]
\vspace{.5cm} $^{(2)}$Physics Department, National Taiwan Normal University,
Taipei, Taiwan, R.O.C. \\[0pt]
\vspace{.5cm} $^{(3)}$Fermilab, P.O. Box 500, Batavia. 
Illinois, 60510, USA.
\\[0pt]
\vspace{.5cm} $^{(4)}$Physics Department, University of Illinois at Chicago,
IL 60607-7059, USA \\[0pt]
}
\end{center}

\begin{abstract}
We study the $B$ meson radiative decay $B \to X_s \gamma$ in the
vector quark model. Deviation from the Standard Model arises from the
non-unitarity of the charged current KM matrix and related new FCNC
interactions.  We establish the relation between the non-unitarity of
charged current mixing matrix and the mixing among the vector quark and
the ordinary quarks.  We also make explicitly the close connection
between this non-unitarity and the flavor changing neutral currents.
The complete calculation including leading logarithmic QCD correction
is carefully carried out.  Using the most updated data and the NLO
theoretical calculation, the branching fraction of the observed $B$
meson radiative decay places a limit on the mixing angles as stringent
as that from the process $B\rightarrow X \mu\bar\mu$.
\end{abstract}

\vspace{1in} \centerline{PACS numbers: 13.38.Dg, 12.15.Ff, 12.90.+b \hfill}
\newpage

\section{Introduction}

A simple extension of Standard Model (SM)  is to enlarge the particle
content by adding vector quarks, whose right-handed and left-handed
components transform in the same way under the weak $SU(2)\times U(1)$
gauge group. This extension is acceptable because the anomalies
generated by the vector quarks cancel automatically and vector quarks
can be heavy naturally.  Vector quarks also arise in some Grand
Unification Theory (GUT). For example, in some superstring theories,
the ${\rm E_6}$ GUT gauge group occurs in four dimensions when we
start with ${\rm E_8
\times E_8}$ in ten dimensions.  The fermions are placed in a
27-dimensional representation of ${\rm E_6}$.  In such model, for each
generation one would have new fermions including an isosinglet charge
$-{1\over3}$ vector quark.

%
%

In this article, we discuss the $B$ meson radiative decay in the
context of a generic vector quark model and show that the experimental
data can
be used to constrain the mixing angles.  In vector quark models, due
to the mixing of vector quarks with ordinary quarks, the
Kobayashi-Maskawa (KM) matrix of the charged current interaction is
not unitary. The internal flavor independent contributions in the $W$
exchange penguin diagrams no longer cancel among the various internal
up-type quarks. In addition, the mixing also generates non-zero tree level 
FCNC in the currents of $Z$ boson and that of Higgs boson, which in turn
gives rise to 
new penguin diagrams due to neutral boson exchanges.  All these  
contributions will be carefully analyzed in this paper.
Leading logarithmic (LL)  QCD corrections are also included
by using the effective Hamiltonian formalism.
The paper is organized as follows: 
In section 2, we review the charged current interaction and the FCNC 
interactions in a generic vector quark model.  Through the diagonalization
of 
mass matrix, the non-unitarity of KM matrix and the magnitude of the FCNC 
can both be related to the mixing angles between vector and ordinary quarks.
In section 3, various contributions to $B$ meson radiative decays are
discussed in the vector quark model.
In section 4, we discuss constraints on the mixing angles from the new 
data on $B$ radiative decays and from other FCNC effects.  
There are many previous analyses on the same issue.  We shall make
detailed comparison at appropriate points (mostly in section 3.) of our
discussion.  Most vector quark models in the literature are more
complicated than the one we considered here.  

\section{Vector Quark Model}
We consider the model in which the gauge structure of SM 
remains while one charge $-{1\over3}$ and one charge 
${2\over3}$ isosinglet vector quarks are introduced. 
Denote the charge $-{1\over3}$ vector quark
as $D$ and the charge ${2\over3}$ vector quark as $U$.
Large Dirac masses of vector quarks, invariant under $SU(2)_L$, 
naturally arise: 
\begin{equation}
M_{\sss U} ( \bar{U}_L U_R + \bar{U}_R U_L) + 
M_{\sss D} (\bar{D}_L D_R + \bar{D}_R D_L)
\end{equation}
All the other Dirac masses can only arise from $SU(2)_L$ symmetry 
breaking effects. Assume that the weak $SU(2)$ gauge 
symmetry breaking sector is an isodoublet scalar Higgs field 
$\phi$, denoted as 
\begin{equation}
\phi \equiv \left( \begin{array}{c}
                   \phi^+ \\
                   \phi^0
                   \end{array}   \right) 
  =   \left( \begin{array}{c}
                   \phi^+ \\
                   \frac{\displaystyle 1}{\displaystyle \sqrt{2}} (v+h^0)
                   \end{array}   \right) 
\end{equation}
We can express the neutral field $h$ in terms of real components:
\begin{equation}
h^0 = H + i \chi.
\end{equation}
The conjugate of $\phi$ is defined as 
\begin{equation}
\tilde{\phi} \equiv \left( \begin{array}{c}
                   \phi^{0*} \\
                   -\phi^-
                   \end{array}   \right) 
  =   \left( \begin{array}{c}
                   \frac{\displaystyle 1}{\displaystyle \sqrt{2}}
                   (v+h^{0*})  \\
                   -\phi^- 
                   \end{array}   \right) 
\end{equation}
Masses for ordinary quarks arise from gauge invariant Yukawa
couplings: 
\begin{equation}
- f_d^{ij} \, \bar{\psi}^i_{\sss L} d^{j}_{\sss R} \phi 
- f_u^{ij} \, \bar{\psi}^i_{\sss L} u^{j}_{\sss R} \tilde{\phi}
- f_d^{ij*} \, \phi^{\dagger} \bar{d}^{j}_{\sss R} \psi^i_{\sss L} 
- f_u^{ij*} \, \tilde{\phi}^{\dagger} \bar{u}^{j}_{\sss R} \psi^i_{\sss L} 
\end{equation} 
In addition, gauge invariant Yukawa couplings between vector 
quarks and ordinary quarks are possible, which give rise to mixing 
between quarks of the same charge. 
For the model we are considering, these are:
\begin{equation}
- f_d^{i4} \, \bar{\psi}^i_{\sss L} D_{\sss R} \phi 
- f_u^{i4} \, \bar{\psi}^i_{\sss L} U_{\sss R} \tilde{\phi}
- f_d^{i4*} \, \phi^{\dagger} \bar{D}_{\sss R} \psi^i_{\sss L} 
- f_u^{i4*} \, \tilde{\phi}^{\dagger} \bar{U}_{\sss R} \psi^i_{\sss L} 
\end{equation}

In general, $U$ will mix with the up-type quarks and $D$ with down-type 
quarks. It is thus convenient to put mixing quarks into a 
four component column matrix:
\begin{equation}
(u_{\sss L,R})_{\alpha} = \left[  \begin{array}{c}
                  u_{\sss L,R} \\ c_{\sss L,R}
                   \\ t_{\sss L,R} \\ U_{\sss L,R} \end{array} \right]_{\alpha}
\  \
(d_{\sss L,R})_{\alpha} = \left[  \begin{array}{c}
                  d_{\sss L,R} \\ s_{\sss L,R} \\ 
                  b_{\sss L,R} \\ D_{\sss L,R} \end{array} \right]_{\alpha}
\end{equation}
where $\alpha=1,2,3,4$. All the Dirac 
mass terms can then be collected into a matrix form: 
\begin{equation}
\bar{d}^{\prime}_{L} {\cal M}_{d} d^{\prime}_{R} +
\bar{d}^{\prime}_{R} {\cal M}_{d}^{\dagger} d^{\prime}_{L} 
\, \, \, \, {\rm and} \, \, \, \,
\bar{u}^{\prime}_{L} {\cal M}_{u} u^{\prime}_{R} +
\bar{u}^{\prime}_{R} {\cal M}_{u}^{\dagger} u^{\prime}_{L}.  
\end{equation}
In this article, we use fields with prime to denote the weak 
eigenstates and those without prime to denote mass eigenstates.
${\cal M}_{u,d}$ are $4 \times 4$ mass matrices.
Since all the right-handed quarks, including vector quark, are 
isosinglet, we can use the right-handed chiral transformation to choose
the right handed quark basis so that $U_L,D_L$ do not have Yukawa coupling
to the ordinary right-handed quarks. In this basis,
${\cal M}_{d}$ and ${\cal M}_u$ can be written as 
\begin{equation}
{\cal M}_{d} = \left( 
\begin{array}{cc}
\hat{M}_{d} & \vec{J}_d \\ 
0 & M_{\sss D}
\end{array}
\right),  \ \ 
{\cal M}_{u} = \left( 
\begin{array}{cc}
\hat{M}_{u} & \vec{J}_u \\ 
0 & M_{\sss U}
\end{array}
\right).
\end{equation}
with
\begin{equation}
\hat{M}_{u,d} = \frac{v}{\sqrt{2}} f_{u,d} , \,\, \,\, 
\vec{J}_{u,d}^i = \frac{v}{\sqrt{2}} f_{u,d}^{i4}
\end{equation}
$\hat{M}_{d,u}$ (with hats) are the standard $3 
\times 3$ mass matrices for ordinary quarks. $\vec{J}_{d,u}$ is the 
three component column matrix which determines the mixings between ordinary 
and vector quarks.
We assume that the bare masses $M_{\sss U,D}$ are much larger $M_W$.
With $M_{\sss U,D}$ factored out, ${\cal M}_{d,u}$ can be expressed in 
terms of small dimensionless parameters $a,b$:
\begin{equation}
{\cal M}_{d} = M_{\sss D} \left( 
\begin{array}{cc}
\hat{a}_d & \vec{b}_d \\ 
0 & 1
\end{array}
\right), \ \ 
{\cal M}_{u} = M_{\sss U} \left( 
\begin{array}{cc}
\hat{a}_u & \vec{b}_u \\ 
0 & 1
\end{array}
\right).
\end{equation}

The mixing matrix $U^{u,d}_L$ of the left-handed quarks and 
the corresponding one $U^{u,d}_R$ for right-handed quarks, defined as,
\begin{equation}
u'_{\sss L,R}= U^{u}_{\sss L,R} u_{\sss L,R}, \ \ 
d'_{\sss L,R}= U^{d}_{\sss L,R} d_{\sss L,R}, 
\end{equation}
are the matrices that diagonalize ${\cal M}_{u,d} {\cal 
M}_{u,d}^{\dagger}$ and
${\cal M}^{\dagger}_{u,d} {\cal M}_{u,d}$ respectively.
Hence the mass matrices can be expressed as
\begin{equation}
{\cal M}_u = U^{u}_{\sss L} m_u  U^{u \dagger}_{\sss R} \,\,\,\,
{\cal M}_d = U^{d}_{\sss L} m_d  U^{d \dagger}_{\sss R} 
\end{equation}
with $m_{u,d}$ the diagonalized mass matrices.
The diagonalization can be carried out order by order in perturbation 
expansion with respect to small numbers $\hat{a}$ and $\vec{b}$. 
For isosinglet vector quark model, the right-handed quark mixings are  
significantly smaller. The reason is that
$M^{\dagger}_{d} M_{d}$ is composed of elements suppressed by two powers 
of $a$ or $b$ except for the $(4,4)$ element.  As a result, the mixings of 
$D_R$ with $d_{R}, s_{R}, b_{R}$ are also suppressed by two powers of $a$ 
or $b$. On the other hand, it 
can be shown that the mixings between $D_L$ and $b_{L},s_{L},d_{L}$ are only 
of first order in  $a$ or $b$.  To get leading order results in the 
perturbation, one can assume that $U_R = I$.
For convenience, write $U_L$ as  
\begin{equation}
U_L = \left( \begin{array}{cc}
\hat{K} & \vec{R} \\ 
\vec{S}^T & T
\end{array}
\right) \;.
\end{equation}
where $\hat{K}$ is a $3 \times 3$ matrix and $\vec{R},\vec{S}$ are three 
component column matrices. 
To leading order in $a$ and $b$, $T$ is equal to $1$. $K$ equals the unitary 
matrix that diagonalizes $\hat{a} \hat{a}^{\dagger}$. The
columns $\vec{R}$ and $\vec{S}$, characterizing the mixing, are given by 
\begin{equation}
\vec{R} = \vec{b}, \ \  \vec{S} = - \hat{K} \vec{b}.
\end{equation}

Now we can write down the various electroweak interactions in terms of
mass eigenstates.
The $Z$ coupling to the left-handed mass eigenstates are given by 
\begin{eqnarray} 
& & {\cal L}_Z = \frac{g}{\cos \theta_W}  Z_{\mu} 
    (J_{3}^{\mu} - \sin^2 \theta_{W} J_{\rm em}^{\mu}), \\
& & J_{3}^{\mu}=\bar{u}'_{L} 
     T^{\rm w}_3 \gamma^{\mu} u'_{L} +
     \bar{d}'_{L} 
     T^{\rm w}_3 \gamma^{\mu} d'_{L} 
    = \frac{1}{2} \bar{u}_{L} (z^u)
      \gamma^{\mu} u_{L}-
      \frac{1}{2} \bar{d}_{L} (z^d)
      \gamma^{\mu} d_{L}
\end{eqnarray}
The $4 \times 4$ matrices $z$ are related to the mixing matrices by
\begin{eqnarray}
(z^{u}) & = & U_L^{u \dagger} a_{\sss Z} U_L^{u} 
\nonumber\\
(z^{d}) & = & U_L^{d \dagger} a_{\sss Z} U_L^{d}. 
\label{FCNC}
\end{eqnarray}
with $a_{\sss Z} \equiv {\rm Diag} (1,1,1,0)$.
Note that the matrix $z$ is not diagonal.
Flavor Changing Neutral Current (FCNC) is generated by the mixings between 
ordinary and vector quarks\cite{lavoura2,branco,silverman}.

The charged current interaction is given by 
\begin{eqnarray}
 &  & {\cal L}_W = \frac{g}{\sqrt{2}} (W_{\mu}^- J^{\mu +} 
    + W_{\mu}^+ J^{\mu -}),  \\
 &  & J^{\mu -}=\bar{u}'_{L} 
     a_{\sss W} \gamma^{\mu} d'_{L} 
    = \bar{u}_{L} V                                                                                                                              
      \gamma^{\mu} d_{L}
\end{eqnarray}
where $a_{\sss W} \equiv {\rm Diag} (1,1,1,a)$ is composed of the 
Clebsch-Gordon coefficients of
the corresponding quarks. For an isosinglet vector quark, 
$a=0$. The $4 \times 4$ generalized KM matrix $V$ is given by: 
\begin{equation}
V= U_L^{u \dagger} a_{\sss W} U_L^d.
\end{equation} 
The standard $3 \times 3$ KM matrix $\VKM$ is the 
the upper-left submatrix of $V$.
Neither $V$ nor $\VKM$ is unitary.
Note that the non-unitarity of $V$ is captured by two matrices
\begin{eqnarray}
(V^{\dagger} V) & = &
U_L^{d \dagger} a^2_{\sss W} U_L^d \nonumber \\
(V V^{\dagger}) & = &
U_L^{u \dagger} a^2_{\sss W} U_L^u. 
\label{KM}
\end{eqnarray}
In the model we are considering, these two matrices are identical to 
$z^{u,d}$ of the FCNC effects in Eq. \ref{FCNC} 
since $a^2_{\sss W}$ is equal to $a_z$. Indeed
\begin{equation}
V^{\dagger} V = (z^d),  \, \, \, \, \, \,
V V^{\dagger} = (z^u)
\end{equation}
This intimate relation between the non-unitarity of $W$ charge current and
the FCNC of $Z$ boson is important for maintaining the gauge
invariance of their combined contributions to any physical process.

The off-diagonal elements of these matrices, characterizing the 
non-unitarity, is closely related to the mixing of ordinary and vector 
quarks. 
The off-diagonal elements are of order $a^2$ or $b^2$. To calculate it, 
in principle, the next-to-leading order expansion of $\hat{K}$, denoted as 
$\hat{K}_2$, is needed. In fact
\begin{equation}
(V^{\dagger} V)_{ij}=(\hat{K}^d_2+\hat{K}_2^{d\dagger})_{ij} + 
a^2 (\vec{b}_d)_i (\vec{b}_d)_j^* 
\end{equation}
Fortunately, by the unitarity of the mixing matrix $U^d$, the 
combination $\hat{K}_2^d+\hat{K}_2^{d\dagger}$ is equal to 
$-(\vec{b}_d)(\vec{b}_d)^{\dagger}$. 
\begin{equation}
\hat{K}_2^d+\hat{K}_2^{d\dagger} = -(\vec{b}_d)(\vec{b}_d)^{\dagger}
\end{equation}
Thus the off-diagonal elements can be simplified
\begin{equation}
(V^{\dagger} V)_{ij}=(-1+a^2) 
(\vec{b}_d)_i (\vec{b}_d)_j^* 
\end{equation}
For isosinglet vector quark, $a=0$.

The Yukawa couplings between Higgs fields and quarks in weak eigenstate 
can be written in a matrix form as 
\begin{equation}
-\frac{v}{\sqrt{2}} \left(
\bar{\psi}'_L a_{\sss Z} {\cal M}_d d'_R \phi 
+\bar{d}'_R {\cal M}_d^{\dagger} a_{\sss Z} \psi'_L \phi^{\dagger} 
+\bar{\psi}'_L a_{\sss Z} {\cal M}_u  u'_R \tilde{\phi} 
+\bar{u}'_R {\cal M}_u^{\dagger} a_{\sss Z} \psi'_L 
\tilde{\phi}^{\dagger} \right)
\end{equation}
Note that $\hat{a}_{\sss Z}$ is added to ensure that the left handed 
isosinglet vector quarks do not participate in the Yukawa couplings. 
The Yukawa interactions of quark mass eigenstates 
and unphysical charged Higgs 
fields $\phi^{\pm}$ are given by
\begin{equation}
{\cal L}_{\phi^{\pm}} = 
\frac{g}{\sqrt{2} M_W} \left[ \bar{u} (m_u V L - 
V m_d R) d \right] \phi^{+} + 
\frac{g}{\sqrt{2} M_W} \left[ \bar{d} (-m_d V^{\dagger} L +
V^{\dagger} m_u R) u \right] \phi^{-}   
\end{equation} 
while those of Higgs boson $H$ and unphysical neutral Higgs field $\chi$ by
\begin{eqnarray}
{\cal L}_{H}  & = &
- \frac{g}{2 M_W} \left[ \bar{d} (m_d z^d L + z^d m_d R )d  +
 \bar{u} (m_u z^u L+z^u m_u R) u \right] H  \\
{\cal L}_{\chi}  & = &
- \frac{ig}{2 M_W} \left[ \bar{d} (- m_d z^d L + z^d m_d R) d  +
 \bar{u} (m_u z^u L- z^u m_u R ) u \right] \chi^0
\end{eqnarray}  

\section{$B$ Meson Radiative Decay}
The $B \to X_s \gamma$ decay, which already exists via one-loop $W$-exchange
diagram in SM, is known to be extremely sensitive to the structure of
fundamental interactions at the electroweak scale and serve as a good
probe of new physics beyond SM because new interaction generically can
also give rise to significant contribution at the one-loop level.

The inclusive $B \to X_s \gamma$ decay is especially interesting. 
In contrast to exclusive decay modes, it is theoretically clean in the
sense 
that no specific low energy hadronic model is needed to describe the
decays. 
As a result of the Heavy Quark Effective Theory (HQET), the
inclusive $B$ meson decay width $\Gamma(B \to X_s \gamma)$ can be well 
approximated by the corresponding $b$ quark decay width
$\Gamma(b \to s \gamma)$. The corrections to this approximation are 
suppressed by $1/m_b^2$ \cite{chay} and is estimated to contribute 
well below $10\%$ \cite{falk,misiak}. 
This numerical limit is supposed to  hold even for the recently discovered
non-perturbative contributions which are
suppressed by $1/m_c^2$ instead of $1/m_b^2$ \cite{voloshin}. 
In the following, we focus on the dominant quark decay $b \to s \gamma$.

In SM, $b\rightarrow s\gamma$ arises at the one loop 
level from the various $W$ mediated penguin diagrams.
The number of diagrams needed to be considered can be reduced by 
choosing the non-linear ${\rm R}_{\xi}$ gauge as in \cite{deshpande}. 
In this gauge, the tri-linear coupling involving photon, $W$ boson and the
unphysical Higgs field $\phi^+$ vanishes. Therefore only four diagrams
, as in Fig.~1. 
contribute. 

\begin{figure}[ht]
\begin{center}
\leavevmode
\epsfbox{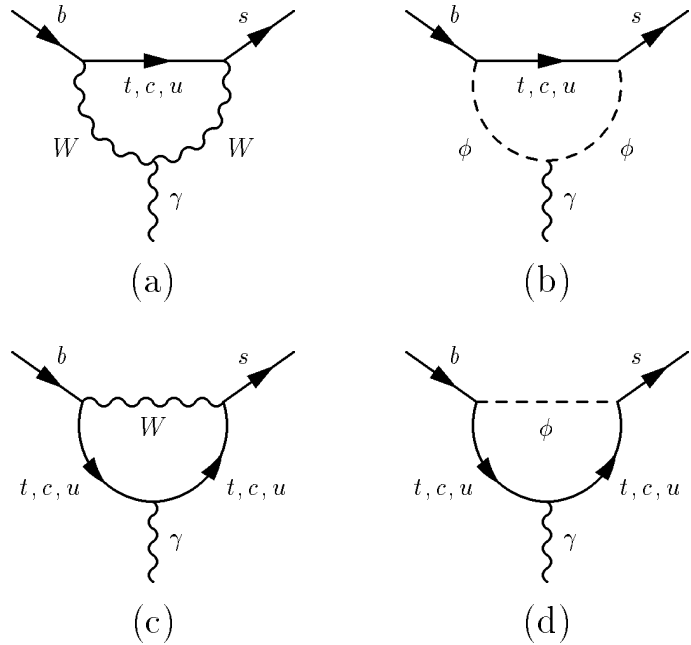}
\caption{Charged boson mediated penguin.}
\end{center}
\end{figure}

The on-shell Feynman amplitude can be written as
\begin{equation}
i {\cal M}(b \rightarrow s \gamma) = \frac{\sqrt{2} G_F}{\pi}
{e\over 4\pi}
\sum_{i} V_{ib} V^*_{is} F_2(x_i) q^{\mu} \epsilon^{\nu} 
\bar{s}\sigma_{\mu \nu}(m_b R+m_s L)b
\end{equation}
with $x_i \equiv m_i^2/M_W^2$ The sum is over the quarks $u,c$ and $t$.
The contributions to $F_2$ from the four diagrams are denoted as 
$f^{W,\phi}_{1,2}$,
with the subscript $1$ used to denote the diagrams with 
photon emitted from internal quark and $2$ those with photon
emitted from $W$ boson.  
The functions $f$'s are given by
\begin{eqnarray}
f^W_1(x)&=& e_i\left[\xi_0(x)-\hbox{$3\over2$}\xi_1(x)
                             +\hbox{$1\over2$}\xi_2(x)
               \right] \ ,\\
f^W_2(x)&=& \xi_{-1}(x)-\hbox{$5\over2$}\xi_0(x)
             +2\xi_1(x)-\hbox{$1\over2$}\xi_2(x) \ ,\\
f^{\phi}_1(x)&=& \hbox{$1\over4$}e_i x \left[ \xi_1(x) + \xi_2(x)\right] \ ,\\
f^{\phi}_2(x)&=& \hbox{$1\over4$} x \left[\xi_0(x)-\xi_2(x)\right]  \ .
\end{eqnarray}
Here the functions $\xi(x)$ are defined as 
\begin{equation}
\xi_n(x) \equiv \int^1_0 \frac{z^{n+1}{\rm d}z}{1+(x-1)z}
=-{ {\rm ln}x+(1-x)+\cdots+{(1-x)^{n+1}\over n+1} \over (1-x)^{n+2} }
\ ,
\end{equation}
and
\begin{equation}
\xi_{-1}(x) \equiv \int^1_0 \frac{{\rm d}z}{1+(x-1)z}
=-{ {\rm ln}x\over 1-x}
\end{equation}
$F_2(x)$ is the sum of these functions and is given by
\begin{equation}  \label{F2}
F_2(x)= f^W_1(x) + f^W_2(x) + f^{\phi}_1(x) +f^{\phi}_2(x) =
\frac{8x^3+5x^2-7x}{24(1-x)^3}-
\frac{x^2(2-3x)}{4(1-x)^4} \ln x + \frac{23}{36}
\end{equation}
For light quarks such as $u$ and $c$, with $x_i \to 0$, the first two 
terms on the right hand side are small and can be ignored. $F_2(x_{u,c})$ is dominated by 
the $x$ independent term ${23\over36}$.
However these mass-independent terms get canceled among the up-type 
quarks due to the unitarity of KM matrix in SM
\begin{equation}
\sum_{i} V_{ib} V^*_{is} = 0 
\end{equation}
After the cancelation, the remaining contributions 
are essentially from penguins with internal $t$ quark.

It is convenient to discuss weak decays using the effective
Hamiltonian formalism \cite{buras,grinstein}, which is crucial for
incorporating the QCD corrections to be discussed later.
In this formalism, the $W$ and $Z$ bosons are integrated out at 
matching boundary $M_{W}$. Their effects are compensated by
non-renormalizable effective Hamiltonian operators.
The important dim-6 operators relevant for $b\rightarrow s \gamma$
 include 12 operators
\begin{equation}
H_{\rm eff}=-\frac{G_F}{\sqrt{2}} V_{ts}^*V_{tb} \sum_{i=1}^{12}
C_i(\mu) O_i
\quad,
\end{equation}
\begin{eqnarray}
Q_1 & = & (\bar{c}_{\alpha}  b_{\beta})_{V-A}
          (\bar{s}_{\beta} c_{\alpha})_{V-A}  
 \nonumber \quad,\\
Q_2 & = & (\bar{c}_{ \alpha}  b_{ \alpha})_{V-A}
          (\bar{s}_{\beta} c_{ \beta})_{V-A}  \nonumber  \\
Q_3 & = & (\bar{s}_{\alpha} b_{\alpha})_{V-A}
          \sum_q (\bar{q}_{ \beta}  q_{\beta})_{V-A} \nonumber\\
Q_4 & = & (\bar{s}_{ \alpha}  b_{ \beta})_{V-A}
          \sum_q (\bar{q}_{ \beta}  q_{ \alpha})_{V-A}\nonumber\\
Q_5 & = & (\bar{s}_{\alpha}  b_{\alpha})_{V-A}
          \sum_q (\bar{q}_{ \beta}  q_{\beta})_{V+A}\nonumber \\
Q_6 & = & (\bar{s}_{\alpha}  b_{ \beta})_{V-A}
          \sum_q (\bar{q}_{ \beta}  q_{ \alpha})_{V+A}\nonumber \\
Q_{7}&=&    \frac{3}{2} (\bar s_{\alpha} b_{ \alpha})_{V-A}  
      \sum_q e_q (\bar q_{\beta}  q_{ \beta} )_{V+A}  
\nonumber\\
Q_{8}&=& \frac{3}{2} (\bar s_{\alpha}  b_{ \beta})_{V-A}  
      \sum_q e_q (\bar q_{\beta}   q_{\alpha})_{V+A}    
\nonumber\\
Q_{9} &=&   \frac{3}{2}    (\bar s_{\alpha} b_{\alpha})_{V-A}  
      \sum_q e_q (\bar q_{ \beta}  q_{ \beta} )_{V-A} 
\nonumber\\
Q_{10} &=&  \frac{3}{2}  (\bar s_{\alpha}  b_{\beta})_{V-A}  
      \sum_q e_q (\bar q_{\beta}  q_{\alpha})_{V-A} \nonumber \\
Q_{\gamma} & = & \frac{e}{8\pi^2}
          \bar{s}_{ \alpha} \sigma^{\mu \nu}
          [m_b (1+\gamma_5)+m_s (1-\gamma_5)]b_{\alpha} 
          F_{\mu\nu}  \nonumber\\
Q_{G} & = & \frac{g_{\rm s}}{8\pi^2}
          \bar{s}_{ \alpha} \sigma^{\mu \nu}
          [m_b (1+\gamma_5)+m_s (1-\gamma_5)] (T_{\alpha\beta}^A) b_{ \alpha} 
          G^A_{\mu\nu}   
\end{eqnarray}
 In Standard Model, the electroweak penguin
operators $Q_{7},\ldots ,Q_{10}$ are not necessary for a leading order
calculation in $O(\alpha)$. However, we will show later that in the vector
quark model, FCNC effects exist as a linear combination of
$Q_7,\ldots,Q_{10}$. This effect will mix with $Q_{\gamma}$
through RG evolution.

The Wilson coefficients $C_i$ at $\mu=M_W$ are determined 
by the matching
conditions when $W$, $Z$ bosons and $t$ quark are integrated out. 
To the zeroth order of $\alpha_s$ and $\alpha$,
the only non-vanishing Wilson coefficients at $\mu=M_W$ for
the above set are $C_{2},C_{\gamma},C_G$.  $C_2$ is given by
\begin{equation}
C_2(M_W) = -V_{cs}^{\ast }V_{cb}/V_{ts}^{\ast }V_{tb}.
\end{equation}
It is equal to one if the KM matrix is unitary and 
$V_{us}^{\ast }V_{ub}$ is ignored.
$C_{\gamma}$ at the scale $M_W$
is given by the earlier penguin calculations
\begin{equation}
C_{\gamma}^{\rm SM}(M_{W})= \frac{1}{V_{ts}^{\ast }V_{tb}}
\sum_{i} V_{ib} V^*_{is} F_2(x_i) =
-\frac{1}{2} D'_0(x_{t})
\simeq -0.193 \quad\quad. 
\end{equation}
The numerical value is given when $m_t=170$ GeV.
Here the function $D'_0$ is
defined as \cite{buras,inami} 
\begin{equation}
D'_0(x) \equiv - \frac{8x^3+5x^2-7x}{12(1-x)^3}+
\frac{x^2(2-3x)}{2(1-x)^4} \ln x.
\end{equation}
$C_{\gamma}$ retains only the mass-dependent contribution from 
the penguin diagram with internal t quark.
The mass-independent terms get cancelled, due to unitarity, among the
three internal up type quarks. The mass-dependent contributions
from the penguin diagrams with internal $u$ and $c$ quarks 
(they are small anyway) appear both 
in the high energy and low energy theories and get canceled in the 
matching procedure. 
Similarly, in SM the $b \to s g$ transition arises from $W$
exchange penguin diagrams which induce $Q_{G}$.
Since the gluons do not couple to $W$ bosons, the gluonic $W$ boson
penguin consists only of two diagrams, which are given by $f^{W,\phi}_1$ 
with $Q$ replaced by one. With the mass-independent contribution 
canceled, the Wilson coefficient $C_G$ can be written as
\begin{equation}
C_{G}^{\rm SM}(M_{W})=-\frac{1}{2} E'_0(x_{t})\simeq 0.096 \quad\quad,
\end{equation}
The function $E'_0$ is defined as \cite{inami}
\begin{equation}
E'_0(x) \equiv - \frac{x(x^2-5x-2)}{4(1-x)^3}+
\frac{3}{2} \frac{x^2}{(1-x)^4} \ln x.
\end{equation}

It is well known that short distance QCD correction is important for 
$b \to s \gamma$ decay and actually enhances the decay rate by more than a 
factor of two. These QCD corrections can be attributed to logarithms of 
the form $\alpha_s^n(m_b) \log^m(m_b/M_W)$. The Leading Logarithmic 
Approximation (LLA) resums the LL series ($m = n$). 
Working to next-to-leading-log (NLL) means that we also resum all the 
terms of the form $\alpha_s(m_b) \alpha_s^n(m_b) \log^n(m_b/M_W)$. 
The QCD corrections 
can be incorporated simply by running the renormalization
scale from the matching scale $\mu=M_W$ down to $m_{b}$ and then calculate 
the Feynman amplitude at the scale $m_b$.
The anomalous dimensions for the RG running 
have been found to be scheme dependent even to LL order, 
depending on how $\gamma_5$ is defined in the dimensional 
regularization scheme.
\chk It has also been noticed \cite{ciuchini} that the one-loop matrix elements of 
$Q_5,Q_6$
for $b \rightarrow s \gamma$ are also regularization scheme dependent. 
The matrix elements of $Q_{5,6}$ vanish in any four dimensional 
regularization scheme and in
the `t Hooft-Veltman (HV) scheme 
but are non-zero in the Naive dimension regularization (NDR)
scheme. This dependence will exactly cancel the
scheme-dependence in the anomalous
dimensions and render a scheme-independent prediction.
We refer to 
\cite{buras,ciuchini} for a review and details.  
In following, we choose to use the HV scheme. 
\footnote{ It is customary in the literature to introduce \cite{pokorski}  
the scheme independent "effective coefficients" for $Q_{\gamma},Q_{G}$.
These coefficients are defined so that $Q_{\gamma},Q_{G}$ are
the only operators with non-zero one loop matrix element for the process
$b \rightarrow s \gamma (g)$.
 The "effective
coeffients" are hence identical to the original
Wilson coefficients
in HV scheme.}

 In the HV scheme, 
only $Q_{\gamma}$ has a non-vanishing matrix 
element between $b$ and $s \gamma$, 
to leading order in $\alpha_s(m_b)$, . 
Thus we only need $C_{\gamma}(m_b)$ to 
calculate the LLA of $b \to s \gamma$ 
decay width. For $m_t=170$ GeV, $m_b=5$ GeV, 
$\alpha_s^{(5)}(M_Z)=0.117$ and in the HV scheme, 
$C_{\gamma}(m_b)$ is related to the non-zero Wilson coefficients at $M_W$ by 
\cite{buras,pokorski,ciuchini} 
\[
C_{\gamma}(m_b)=0.698\,C_{\gamma}(M_{W})+0.086\,C_{G}(M_{W})
-0.156\,C_{2}(M_{W}).
\]
The $b \to s \gamma$ amplitude is given by
\begin{equation}
{\cal M}(b \to s \gamma) = - V_{tb} V^{*}_{ts} {G_F \over \sqrt{2}} 
C_{\gamma}(m_b) \langle Q_{\gamma} \rangle_{\rm tree}
\end{equation}
To avoid the uncertainty in $m_b$, it is 
customary to calculate the ratio $R$ between the radiative decay and the 
dominant semileptonic decay.
The ratio $R$ is given, to LLA, by \cite{pokorski}
\begin{equation}
R \equiv \frac{\Gamma(b \rightarrow s \gamma)}
{\Gamma(b \rightarrow c e \bar{\nu}_{e})} 
=\frac{1}{\left| V_{cb}\right|^{2}}\frac{6\alpha}{\pi g(z)}
\left| V_{ts}^{\ast}V_{tb}C_{\gamma}(m_b)\right|^{2}. \label{RR}
\end{equation}
Here the function $g(z)$ of $z=m_c/m_b$ is defined as 
\begin{equation}
g(z)=1-8z^2+8z^6-z^8-24z^4 \ln z
\end{equation}

In the vector quark model, deviations from SM result
come from various sources: (1) charged current KM matrix
non-unitarity, (2) Flavor Changing Neutral Current (FCNC) effects in
neutral boson mediated penguin diagrams, and (3) the $W$ penguin with
internal heavy $U$ vector quark. Since the last one can be
incorporated quite straight-forwardly, we do not elaborate on this
contribution which will not be relevant for models without the $U$
quark. 
We concentrate on the first two contributions, which have been
discussed in Refs.\cite{brancobsg,handoko,barger}.  Here we make a more
careful and complete analysis which supplements or corrects these earlier
analyses. 
Refs.\cite{brancobsg} 
have calculated  effects due to  non-unitarity of the KM
matrix and effects  due to the $Z$  mediated penguin in the Feynman
gauge, however, their analysis did not
include the FCNC contribution from the unphysical neutral Higgs boson, 
which is necessary for gauge invariance.  The Higgs boson mediated 
penguins were also ignored.  On the other hand,
Ref.\cite{handoko}, while taking the unphysical Higgs boson into account,
did not consider effects due to non-unitarity of the KM matrix,
which gives the most important contribution.  
None of the above treatments, except Ref.\cite{barger},
included QCD corrections.

Our strategy is first to integrate out the vector quark together with $W$ and
$Z$ bosons at the scale $M_W$. 
As shown above, the KM matrix is not unitary in the
presence of an isosinglet vector quark.
Therefore the mass-independent contributions in Eq. (\ref{F2})
from the magnetic penguin diagrams with various up-type 
quarks no longer cancel.
This contribution
is related to the short distance part of loop integration, i.e.\  when the
loop momenta are large so that
the quark mass which appears in the propagator can be ignored.
In the formalism of the low energy effective Hamiltonian, 
it can be shown that
these mass independent contributions never arise if the theory is renormalized
using DRMS or similar schemes. 
By dimensional analysis, it is clear that the corresponding
diagrams calculated using effective Hamiltonians are always proportional
to the square of the loop quark mass.
When we match the two calculations at the $M_{W}$ scale, the
mass-independent contributions
must be compensated by new terms in Wilson coefficients.
This is consistent with the notion that the effective field theory formalism 
separates the short
distance physics encoded in Wilson coefficients from
the long distance physics parameterized by  matrix elements of the effective
Hamiltonian. Such separation enables us to calculate  effects of 
new physics by simply calculating  Wilson coefficients perturbatively
at the matching boundary. The matching results serve as
initial conditions when Wilson coefficients
run to a lower scale by renormalization group. 
Since the vector quarks have been integrated out, the anomalous dimensions 
are not affected by the new physics. 

Following this procedure, we calculate the extra contributions to the
Wilson coefficient $C_{\gamma}$ from non-unitarity:
 \begin{equation}
\frac{(V^{\dagger} V)_{23}}
{V_{ts}^{*}V_{tb}} \, \frac{23}{36} 
=\frac{\delta}{V_{ts}^{*}V_{tb}} \, \frac{23}{36} \ .
\label{eq:nonu}
\end{equation}
The parameter $\delta$, one of the off-diagonal elements of the matrix 
$V^{\dagger} V$, characterizes the
non-unitarity:  
\begin{equation}
\delta = (V^{\dagger} V)_{23}=z_{sb}
\label{eq:delta_z}
\end{equation}

The $b \rightarrow s \gamma$ transitions also arise from FCNC $Z$ 
boson and Higgs boson mediated penguin diagrams as in Fig.~2. 
The FCNC contribution to $C_{\gamma}(M_W)$ can be denoted as follows:
\begin{equation}
\frac{z_{sb}}{V_{tb}V^*_{ts}}(f_{s,b}^Z+f_{s,b}^{\chi}+f_{s,b}^{H})+
\frac{z_{4b}z^{\ast}_{4s}}{V_{tb}V^*_{ts}}(f_D^Z+f_D^{\chi}+f_D^{H})
\end{equation}
For the sake of gauge invariance, $f^Z$ needs to be considered together
with 
$f^{\chi}$.
The $Z$ boson penguins consist of internal charge $-{1\over3}$ quarks.
The contribution from internal $i=b,s$ quark, $f^Z_{s,b}$, is given by  
($y_i \equiv m_i^2/M_Z^2$): 
\begin{eqnarray}
f^Z_b &=& -\hbox{$1\over2$} e_d \left\{ (-\hbox{$1\over2$}-e_d \sin^2 \theta_W)
\left[\,2\xi_0(y_b)-3\xi_1(y_b)+\xi_2(y_b)\,\right]  \right. \nonumber \\
& & \left.\,\,\,\,\,\,\,\, + \, e_d \sin^2 \theta_W  
\left[\,4\xi_0(y_b)-4\xi_1(y_b)\,\right] \right\}  \\ 
f^Z_s &=& -\hbox{$1\over2$} e_d
\left\{ ( -\hbox{$1\over2$}-e_d \sin^2 \theta_W)
\left[\,2\xi_0(y_s)-3\xi_1(y_s)+\xi_2(y_s)\,\right]  \right. \nonumber \\
& & \left.\,\,\,\,\,\,\,\, + \, \frac{m_s}{m_b} e_d \sin^2 \theta_W  
\left[\,4\xi_0(y_b)-4\xi_1(y_b)\,\right]  
 \right\}  
\end{eqnarray}
The last term in $f^Z_s$ has a mass insertion in the
$s$ quark line. It is suppressed by $m_s/m_b$ and will be ignored.  
\begin{figure}[ht]
\begin{center}
\leavevmode
\epsfbox{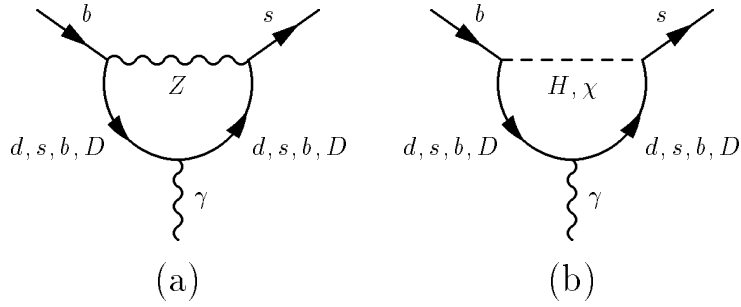}
\caption{Neutral boson mediated penguin diagrams.}
\end{center}
\end{figure}
The  calculation is similar to that of $f^W$. 
For a consistent approximation, the two variables $y_b$ and $y_s$, 
which are the ratios of $m_b^2,m_s^2$ to $M_Z^2$, are also set to zero. 
Hence
\begin{eqnarray}
f^Z_b+f^Z_s &\approx& -\hbox{$1\over2$} {e_d}  
\left\{ (-\hbox{$1\over2$} - e_d \sin^2 \theta_W)
\left[\,4\xi_0(0)-6\xi_1(0)+2\xi_2(0)\,\right] \right. \nonumber \\
& & \left. \,\,\,\,\,\,\,\, + e_d \sin^2 \theta_W  
\left[\,4\xi_0(0)-4\xi_1(0)\,\right] 
\right\}   \\ &=& -\frac{1}{9}-\frac{1}{27} \sin^2 \theta_W \simeq  -0.12
\end{eqnarray} 
The $Z$-mediated penguin diagram with internal $D$ quark can also be 
calculated.
\begin{equation}
f^Z_D  =  {1\over4} e_d 
\left[\,2\xi_0(y_D)-3\xi_1(y_D)+\xi_2(y_D)\,\right]  
\rightarrow  -{5\over 72 y_D}  +O(\frac{1}{y_D^2}) \ . 
\end{equation}
It approaches zero when $y_D \rightarrow \infty$ and thus
$f^Z_D$ is negligible in large $y_D$ limit.
For a gauge invariant result, the unphysical neutral Higgs $\chi$ mediated 
penguin needs to be considered together with the $Z$ boson penguin.
In the non-linear Feynman gauge we have chosen, the mass of $\chi$ is 
equal to $M_Z$. The calculation is very similar to the $\phi^{\pm}$ penguin.
For internal $s,b,D$ quarks, the contributions
$f^{\chi}_{s,b,D}$ are given by
\begin{eqnarray}
f^{\chi}_{i} &=&   \frac{e_d}{8} 
y_{i} \left[\,\xi_1(y_{i})+\xi_2(y_{i})\,\right] \nonumber \\
& = &  - \frac{e_d}{8} y_i \left[(2-y_i)\ln y_i - 
\frac{5}{6} y_i^3+ 4 y_i^2- \frac{13}{2} y_i+\frac{10}{3} \right]
\frac{1}{(1-y_i)^4} 
\end{eqnarray}
It is obvious that the light quark contributions are suppressed by the
light quark masses and thus negligible.
The situation is quite different for the heavy $D$ quark.
As an approximation, for $y_D \to \infty$, $f^{\chi}_D \to -{5\over144} \sim 
-0.035$. 
This contribution, comparable to the $Z$ mediated penguin $f^Z$ from 
light quarks, has been $overlooked$ in previous 
calculations\cite{brancobsg}.
Since the quark $D$ may not be much heavier than $Z$ boson, we expand 
$f^{\chi}_D$ in powers of $1/y_D$ and keep also the next leading term. 
\begin{equation}
f^{\chi}_D \approx - \frac{5}{144} + \frac{1}{36} \frac{1}{y_D} + 
O(\frac{1}{y_D^2}) \ .
\end{equation}

The Higgs boson $H$ mediated penguin is similar to that of unphysical 
Higgs $\chi$:
\begin{eqnarray}
f^{H}_{i} &=& - \frac{e_d}{8} 
w_{i} \left[3 \xi_1(w_{i})-\xi_2(w_{i})\right] \nonumber \\
& = & - \frac{e_d}{8} w_i \left[(-2+3w_i)\ln w_i 
+\frac{7}{6} w_i^3-6 w_i^2+ \frac{15}{2}w_i-\frac{8}{3} \right]
\frac{1}{(1-w_i)^4}
\end{eqnarray} 
where $w_i \equiv m_i^2 / M_H^2$. Similar to the $\chi$ penguin, 
$f^H_{s,b}$ can be ignored since $m_s,m_b \ll m_H$. 
For $f^H_D$, we again expand it in powers of $1/w_D$ and keep up to the 
next leading term: 
\begin{equation}
f^H_D \approx + \frac{7}{144} - \frac{1}{18} \frac{1}{w_D} + 
O(\frac{1}{w_D^2}) 
\end{equation}
The leading term is $+0.048$, again comparable to the $Z$ penguin. 

Put together, the Wilson coefficient $C_{\gamma}(M_W)$ in the vector quark model
is given by 
\begin{eqnarray}
&&\quad\quad C_{\gamma}(M_W) =
C_{\gamma}^{\rm SM}(M_W)  + \frac{\delta}{V_{tb}V^*_{ts}}\frac{23}{36}
+\frac{z_{sb}}{V_{tb}V^*_{ts}}(f_s^Z+f_s^{\chi}+f_s^{H}
                              +f_b^Z+f_b^{\chi}+f_b^{H})   
\nonumber\\
&& \hskip 3in
+\frac{z_{4b}z_{4s}^*}{V_{tb}V^*_{ts}}
                               (f_D^Z+f_D^{\chi}+f_D^{H})  
\nonumber\\
&& = C_{\gamma}^{\rm SM}(M_W) 
+ \frac{z_{sb}}{V_{tb}V^*_{ts}} \, \left( \frac{23}{36} - 
\frac{1}{9} - \frac{1}{27} \sin^2  \theta_W 
+ {5\over 72 y_D}  + \frac{5}{144} - \frac{1}{36} \frac{1}{y_D}
- \frac{7}{144} +\frac{1}{18} \frac{1}{w_D} \right) 
\nonumber\\
&&
\rightarrow
-0.193 + \frac{z_{sb}}{V_{tb}V^*_{ts}} \times 0.506      \quad .
\end{eqnarray}
Here we have used the unitarity relations 
$z_{4b} z^*_{4s} = -|U_{44}|^2 z_{sb} \approx -z_{sb}$ to leading 
order in FCNC due to the unitarity of $U_L^d$ and 
$\delta=z_{sb}$ from Eq. (\ref{eq:delta_z}).
In the above numerical estimate we took $y_D$, $w_D$ to infinity.

Similarly the Wilson coefficient of the gluonic magnetic-penguin operator 
$Q_{G}$ is modified by the vector quark. 
In the vector quark model, the mass-independent term will give an extra 
contribution ${1\over3}  \delta$  if the KM matrix is 
non-unitary\cite{deshpande}.
The FCNC neutral boson mediated gluonic magnetic penguin diagrams are 
identical to those of the photonic magnetic penguin, except for a trivial 
replacement of $Q_d$ by  color factors, since photon and gluons do
not 
couple to neutral bosons. 
$C_G(M_W)$ in the vector quark model is given by
\begin{eqnarray}
&&\quad\quad C_G(M_W)
=  C_G^{\rm SM}(M_W)  + \frac{\delta}{V_{tb}V^*_{ts}}\frac{1}{3}
-3\frac{z_{sb}}{V_{tb}V^*_{ts}}(f_s^Z+f_s^{\chi}+f_s^{H}
                              +f_b^Z+f_b^{\chi}+f_b^{H})   
\nonumber\\
&& \hskip 3in
-3\frac{z_{4b}z_{4s}^*}{V_{tb}V^*_{ts}}
                               (f_D^Z+f_D^{\chi}+f_D^{H})  
\nonumber\\
&&
=  C_G^{\rm SM}(M_W) + \frac{z_{sb}}{V_{tb}V^*_{ts}} \left( \frac{1}{3} + 
\frac{1}{3} + \frac{1}{9} \sin^2 \theta_W 
-{5\over 24 y_D}
- \frac{5}{48} + \frac{1}{12} \frac{1}{y_D}
+ \frac{7}{48} - \frac{1}{6}  \frac{1}{w_D} \right)   
\nonumber\\
&&
  \to  -0.096 + \frac{z_{sb}}{V_{tb}V^*_{ts}} \times 0.733 \quad .
\end{eqnarray}

The above deviation from SM does not include QCD evolution. We can 
incorporate LL QCD corrections to these deviations in the 
framework of effective Hamiltonian.
The key is that the deviation from Standard Model is a short distance 
effect at the scale of $M_W$ and $M_Q$. It can be separated into the 
Wilson coefficients at the matching scale, as we just did. The 
evolution of Wilson coefficients, which incorporates the LL QCD 
corrections, is not affected by the short distance 
physics of vector quark model 
and all the anomalous dimensions used in SM
calculation still valid here.  One only needs to use the corrected Wilson 
coefficients at $\mu=M_W$ and in so doing we resum all the terms of the 
form $z_{sb} \alpha_s^n(m_b) \log^n(m_b/M_W)$. 

One subtlety in the vector quark model is that the quark mixing will generate 
Flavor Changing Neutral Current that couple to $Z$ boson, which
in turn gives rise to $Z$ boson exchange interaction. This interaction 
is represented by an Effective Hamiltonian which is
a linear combination of strong penguin operator $Q_3$ and 
electroweak penguin operator $Q_{7,9}$:
\begin{eqnarray}
H_{\rm NC}&= & 2 \frac{G_F}{\sqrt{2}} 
z_{sb} \left(-{1\over2}\right) (\bar s_{\alpha} b_{ \alpha})_{V-A}  
      \sum_q (t_3-e_q \sin^2 \theta_W) (\bar q_{\beta}  q_{ \beta} )_{V \pm A} \nonumber \\
 & = & -\frac{G_F}{\sqrt{2}} z_{sb}  \left[ -{1\over6} Q_3 -
 {2\over3}\sin^2 \theta_W Q_7 +{2\over3} (1-\sin^2 \theta_W) Q_9 \right]
\end{eqnarray} 
which gives additional non-zero Wilson coefficients:
\begin{eqnarray}
C_3(M_W)&=&-\frac{z_{sb}}{V_{tb}V^*_{ts}}{1\over6} \ ,\nonumber\\
C_7(M_W)&=&-\frac{z_{sb}}{V_{tb}V^*_{ts}}{2\over3}\sin^2 \theta_W \ ,
                                                      \nonumber\\
C_9(M_W)&=&\frac{z_{sb}}{V_{tb}V^*_{ts}}{2\over3}(1-\sin^2 \theta_W) \ .
\end{eqnarray}
To LL, the strong penguin and electroweak penguin operators could
mixing among themselves and also with $Q_{\gamma}$ and $Q_{G}$. 
This will generate an additional 
LL QCD correction to $b \rightarrow s \gamma$ decay in
the vector quark model. The crucial Wilson coefficient $C_{\gamma}(m_b)$ 
obtain additional contributions:
\begin{eqnarray}
C_{\gamma}(m_b) & = & 0.698\,C_{\gamma}(M_{W})+0.086\,C_{G}(M_{W})
-0.156\,C_{2}(M_{W}) \nonumber  \\
& & +0.143\,C_{3}(M_{W})+0.101\,C_{7}(M_{W})
-0.036\,C_{9}(M_{W}).
\end{eqnarray}
This FCNC LL QCD corrections is about one fifth the FCNC contribution of Z boson
mediated penguin. 
The detail of the RG running calculation is given in the appendix.

The correction to ratio $R$ in the vector quark model, including its 
LL QCD corrections, is given by 
\[
\Delta R=\frac{6\alpha}{\pi g(z) \, |V_{cb}|^2 }\times 0.307 \times {\rm Re}
\left[ V_{ts}^{\ast}V_{tb} \, z_{sb} \right] = 
0.23 \, {\rm Re} \, z_{sb}
\] 
to leading order in $\delta$.
In this result, the difference between $V_{ts}^{*} V_{tb}$ and 
$- V_{cs}^{*} V_{cb}$, i.e.
\begin{equation}
 V_{cs}^{*} V_{cb} = z_{sb} -V_{ts}^{*} V_{tb}, 
\end{equation}
has been taken into account.
We use the value $|V_{ts}^{*} V_{tb}|^2/|V_{cb}|^2=0.95$.
 
\section{Constraints}
The inclusive $B \to X_s \gamma$ branching ratio has been measured by CLEO 
with the branching ratio 
\cite{cleo,cleo1} 
\begin{equation}
{\cal B} (B \to X_s \gamma )_{\rm EXP}=(3.15\pm 0.54)\times 10^{-4}
\end{equation}
This branching ratio could be used to constrain the mixing in 
the vector quark model. 
We calculate the vector quark model deviation to leading logarithmic 
order, ie. all the terms of the form 
$z_{sb} \alpha_s^n(m_b) \log^n(m_b/M_W)$.
The Standard Model prediction to leading logarithmic order is \cite{reina}
\begin{equation}
{\cal B}(B \to X_s \gamma )_{\rm LO}=(2.93 \pm 0.67)\times 10^{-4}
\end{equation}
The difference between the
experimental data and the Standard Model LO prediction,
\begin{equation}
{\cal B}(B \to X_s \gamma )_{\rm EXP}-
{\cal B}(B \to X_s \gamma )_{\rm NLO}=(0.22 \pm 0.86)\times 10^{-4}
\end{equation}
It gives a range of possible vector quark model deviation and hence 
on $z_{sb}$ (with the input ${\rm B}(B \rightarrow X_c e
\bar{\nu})=0.105$): 
\begin{equation}
-0.0027<  z_{sb} < 0.0045  
\end{equation}

\chk The SM prediction up to next-to-leading order has been
calculated in Ref.\cite{misiak}, with the result 
\begin{equation}
{\cal B}(B \to X_s \gamma )_{\rm NLO}=(3.28\pm 0.33)\times 10^{-4}
\end{equation}
Ref.\cite{pott} later did a new analysis, which discards all 
corrections beyond NLO by expanding formulas like Eq.(\ref{RR}) in powers of
$\alpha_s$, and reported a slightly higher result: 
\begin{equation}
{\cal B}(B \to X_s \gamma )_{\rm NLO}=(3.60\pm 0.33)\times 10^{-4}
\end{equation}
Here we also use these new
next-to-leading order SM calculations and the leading order
vector quark model correction to
constraint $Z_{sb}$.
To be consistent in the estimate of the theoretical errors, however,
a full next-to-leading order calculation of
the vector quark model matching correction is required.
The difference between the
experimental data and the Standard Model NLO prediction,
with the errors added up directly, is
\begin{eqnarray}
{\cal B}(B \to X_s \gamma )_{\rm EXP}-
{\cal B}(B \to X_s \gamma )_{\rm NLO}&=&(-0.13 \pm 0.63)\times 10^{-4}
 \,\,\,\,\,  {\rm \cite{misiak}} \nonumber \\
                                       & &(-0.45 \pm 0.63) \times 10^{-4}
 \,\,\,\,\,  {\rm \cite{pott}} 
\end{eqnarray}
It gives a constraint
on $z_{sb}$:
\begin{eqnarray}
-0.0032< & z_{sb} <& 0.0021  \,\,\,\,\,  {\rm \cite{misiak}} \nonumber \\
-0.0045< & z_{sb} <& 0.0007 \,\,\,\,\,  {\rm \cite{pott}} 
\end{eqnarray}

The previously strongest bound on $z_{sb}$ is from 
$Z$-mediated FCNC effect in the mode $B\rightarrow X\mu^{+}\mu^{-}$ 
\cite{branco}: \begin{equation}
-0.0012< z_{sb} <0.0012   \label{FCNCB}
\end{equation}
Our new bound is as strong as that from FCNC. 
It shows that even though the vector quarks contribute to the
radiative decay rate through one loop, as in SM, the data could still put
strong bound.

On the other hand, in models like Ref.~\cite{cck}, 
operators of different chiralities such as
\begin{eqnarray}
{O'}_{\gamma} = \frac{e}{8 \pi^2}
          \bar{s}_{\sss \alpha} \sigma^{\mu \nu}
          [m_b (1-\gamma_5)+m_s (1+\gamma_5)]
           b_{\sss \alpha} F_{\mu\nu} \ ,  \nonumber\\
{O'}_{G} = \frac{g_{\rm s}}{8 \pi^2}
          \bar{s}_{\sss \alpha} \sigma^{\mu \nu}
          [m_b (1-\gamma_5)+m_s (1+\gamma_5)]
          (T_{\sss \alpha\beta}^A) b_{\sss \alpha} 
          G^A_{\mu\nu} \ ,  
\end{eqnarray}
occurs via the new interaction.  
Our study can be extended to these models too. However, the new
amplitude for $b\to s \gamma$ belongs to a different helicity 
configuration in the final state and it
will not interfere with the SM contribution. Consequently, 
the constraint obtained from $b\to s\gamma$ in these models is
less stringent than that from $B\to X\mu^+\mu^-$.

In the upcoming years, much more precise measurements
are expected from the upgraded CLEO detector, as well as from the
$B$-factories presently under construction at SLAC and KEK.  The new
experimental result will certainly give us clearer evidence whether the
vector quark model is viable.

DC wishes to thank T. Morozumi and E. Ma for discussions.  WYK is
supported by a grant from DOE of USA.  DC and CHC are supported by grants
from NSC of ROC.
CHC's support by NSC is under contract No.NSC 88-2112-M-003-017. 

\section*{Appendix}
\def\barbarf{{ \overline{\overline{f}} }}

The RG equations for the Wilson coefficients
$\crb \equiv (C_1(\mu), \ldots, C_{10}(\mu))$ and $C_{\gamma}$, $C_G$
can be written as \cite{ciuchini}  
\begin{eqnarray}
\frac{d}{d \ln \mu} \crb(\mu)& = &\frac{\alpha_s}{4\pi} \hat{\gamma}_r^{\rm T} \crb(\mu) 
\nonumber\\
\frac{d}{d \ln \mu} C_{\gamma}(\mu)& = &\frac{\alpha_s}{4\pi}
( \betab_{\gamma} \cdot \crb(\mu)+\gamma_{\gamma\gamma} C_{\gamma}(\mu)
+\gamma_{G \gamma} C_{G}(\mu) )  \nonumber  \\
\frac{d}{d \ln \mu} C_{G}(\mu)& = &\frac{\alpha_s}{4\pi}
( \betab_{G} \cdot \crb(\mu)
+\gamma_{G G} C_{G}(\mu)  ) 
\end{eqnarray}
The 10 by 10 submatrix $\hat{\gamma}_{r}$ can be found in \cite{ciuchini}.  
The anomalous dimension matrix entries $\betab_{\gamma,G}^{7-10}$ 
are extracted from the results of $\betab_{\gamma,G}^{3-6}$ in 
Ref.\ \cite{ciuchini}. \chk In the HV scheme, $\betab_{\gamma,G}$ are given by: 
      \begin{equation}
      \betab_{\gamma} = \left(
             \begin{array}{c}
             0 \\
             -\frac{4}{27} \CA - \frac{2 e_u}{e_d} \CA  \\
             -{116\over27} \CA \\
             -{4 f\over27}\CA - {2 {\bar f}}\CA    \\
             {8\over 3} \CA              \\
             -{4 f\over27}\CA + {2 {\bar f}}\CA  \\
             4 e_d \CA                \\
             -{2 \bar f\over9}e_d\CA  + 3 \barbarf   \CA  \\
             -{58\over9} e_d \CA \\
             -{2 \bar f\over9}e_d \CA - 3 \barbarf   \CA   
              \end{array}
             \right),  
      \,\,\,\,\,\,\,\,\,
      \betab_{G} = \left(
             \begin{array}{c}
             3 \\
             \frac{11N}{9} - \frac{29}{9N}  \\
             {22N\over9}-{58\over 9N}+3f \\
             6  + {11N f\over 9} - {29f\over9N}    \\
             -2N +{4\over N} - 3f             \\
             -4 - {16Nf\over 9} +{25 f \over 9N}   \\
             -3N e_d+{6\over N} e_d- {9\bar f \over 2} e_d             \\
             -6 e_d- {8N {\bar f} \over 3}e_d +{25 {\bar f} \over 6N}e_d  \\   
            {11N\over3}e_d-{29\over 3N}e_d+{9\bar f\over 2}e_d \\
             9 e_d + {11N {\bar f}\over 6} e_d- {29{\bar f}\over6N} e_d                 
             \end{array}
             \right)  
      \end{equation}
Here  $u$ and $d$ are the numbers of active up-type quarks and down type quarks
respectively, $f$ is the total number of active quark flavor. 
Between the scales $m_b$ and $M_W$, $u=2$, $d=3$, $f=5$,
${\bar f} \equiv (e_d d + e_u u)/e_d=-1$,
$\barbarf  \equiv (e_d^2 d + e_u^2 u)/e_d=-11/3$.
For $SU(3)$ color, $N=3$ .

\end{document}